\documentstyle[epsfig]{aipproc2}
\begin{document}

\title{Implementing fully relativistic hydrodynamics
	in three dimensions}
\author{T. W. Baumgarte$^{1}$, S. A. Hughes$^{1}$, L. Rezzolla$^{1}$, 
	S. L. Shapiro$^{1,2}$ and 
	M. Shibata$^{1,3}$}
\address{$^{1}$Department of Physics, 
	University of Illinois at Urbana-Champaign, Urbana, Il 61801 \\
	$^{2}$Department of Astronomy and NCSA,
	University of Illinois at Urbana-Champaign, Urbana, Il 61801 \\
	$^{3}$Department of Earth and Space Science,
	Osaka University, Toyonaka, Osaka 560-0043, Japan}

\maketitle

\begin{abstract}
We report on our numerical implementation of fully relativistic
hydrodynamics coupled to Einstein's field equations in three spatial
dimensions.  We briefly review several steps in our code development,
including our recasting of Einstein's equations and several tests
which demonstrate its advantages for numerical integrations.  We
outline our implementation of relativistic hydrodynamics, and present
numerical results for the evolution of both stable and unstable
Oppenheimer-Volkov equilibrium stars, which represent a very promising
first test of our code.
\end{abstract}

\section*{Introduction}

The physics of compact objects is entering a particularly
exciting phase.  New instruments, including X-Ray and Gamma-Ray
satellites and the new neutrino observatories SNO and
Super-Kamiokande, can now yield unprecedented observations of neutron
stars and black holes.  Perhaps most excitingly, the new gravitational
wave detectors LIGO, TAMA, GEO and VIRGO promise to open a
gravitational wave window to the Universe and make gravitational wave
astronomy a reality (see, e.g.,~\cite{twb:ps99}).

Simultaneously, the availability of computational resources at modern
supercomputers makes the simulation of realistic astrophysical
scenarios involving relativistic compact objects feasable.  Several
groups, including two ``Grand Challenge Alliances''~\cite{twb:gc},
have launched efforts to construct numerical codes capable of solving
Einstein's equations with or without matter sources in three
dimensions, and simulating the merger of black hole or neutron star
binaries (see, e.g.,
\cite{twb:on89,twb:on97,twb:wmm96,twb:bcsst97,twb:bgm99,twb:fmst98,twb:s99}).
Numerical simulations will be necessary to predict the gravitational
wave form from such processes to increase the likelihood
of detections and, ultimately, to extract physical information from
observations.  Even though much progress has recently been made
(see~\cite{twb:s99} for the most recent developments), significant
obstacles still remain.

In this contribution, we report on our systematic approach towards
constructing such a numerical code.  We first review our formulation
of Einstein's equations and describe several tests, both for vacuum
spacetimes and analytical matter sources, which demonstrate its
advantages for numerical integrations.  We then outline our
implementation of hydrodynamics, and present promising numerical test
results.  We adopt geometrized units (G = c = 1) and the convention
that Greek indices run from 0 to 3, while Latin indices only run from
1 to 3.

\section*{Evolution of the Gravitational Fields}

Most numerical implementations of Einstein's equations adopt a Cauchy
formulation based on the 3+1 formulation by Arnowitt, Deser and Misner
(\cite{twb:adm62}, see, e.g.,~\cite{twb:gw92} for an alternative
characteristic formulation).  However, a straightforward
implementation of these ``undressed'' ADM equations tends to develop
instabilities even for the evolution of small amplitude gravitational
waves on a flat background (compare~\cite{twb:aetal98}).  Following
Shibata and Nakamura~\cite{twb:sn95} and the spirit of many earlier
one and two-dimensional codes~\cite{twb:e84}, we have recently
developed a modification of the ADM equations which proves to be much
more suitable for numerical implementations~\cite{twb:bs99}.

More specifically, we modify the ADM equations in two ways.  First,
we split the spatial metric $\gamma_{ij}$ into a conformal factor
$\exp(\phi)$ and a conformally related metric $\tilde \gamma_{ij}$
according to
\begin{equation}
\gamma_{ij} = e^{4 \phi} \tilde \gamma_{ij}
\end{equation}
and evolve $\phi$ and $\tilde \gamma_{ij}$ separately.  We choose
$\phi$ such that $\det(\tilde \gamma_{ij}) = 1$, and similarly split
the extrinsic curvature into its trace and its trace-free part.  This
split separates the ``radiative'' variables from the ``non-radiative''
variables in the spirit of the ``York-Lichnerowicz''
decomposition~\cite{twb:yl}.

In the second stage we introduce ``conformal connection functions''
\begin{equation}
\tilde \Gamma^i \equiv \tilde \gamma^{lm} \tilde \Gamma^i_{lm} 
	= - \tilde \gamma^{il}_{~~,l}
\end{equation}
as independent functions (compare~\cite{twb:nok87}).  Some of
the mixed second derivatives of $\tilde \gamma_{ij}$ in the conformal,
spatial Ricci tensor $\tilde R_{ij}$ can now be written as first
derivatives of the $\tilde \Gamma^i$.  As a result, $\tilde R_{ij}$
becomes a manifestly elliptic operator on the metric $\tilde
\gamma_{ij}$.  The analogous technique was used for the
four-dimensional Ricci tensor $R_{\alpha\beta}$ as early as in the
1920's to make Einstein's equations manifestly
hyperbolic~\cite{twb:c62}.  For more details of our
formulation, including an evolution equation for the $\tilde
\Gamma^i$, the reader is referred to~\cite{twb:bs99} (see also
the mathematical analysis in~\cite{twb:fr99}, and further
numerical applications in~\cite{twb:aablsst99}).

In~\cite{twb:bs99}, we tested this form of Einstein's equations for
small amplitude gravitational waves, and found that it performs far
better than a similar implementation of the original ADM formulation.
In particular, we found that we could evolve such waves with harmonic
slicing without encountering growing instabilities, whereas the
original ADM code crashed after about 35 light-crossing times for the
same initial data.  In~\cite{twb:bhs99}, we inserted analytic matter
sources on the right hand side of Einstein's equations and evolved the
fields in their presence.  This approach allows us to study the
numerical properties of the field evolution in the presence of highly
relativistic matter sources without having to solve the equations of
hydrodynamics: ``hydro-without-hydro''.  We inserted the
Oppenheimer-Snyder solution for a relativistic, static star to test
the long term stability of the evolution code, and the
Oppenheimer-Volkov solution for the collapse of a sphere of dust to a
Schwarzschild black hole.  These simulations focus on the highly
relativistic, longitudinal fields, and complement our earlier tests
involving dynamical transverse fields in~\cite{twb:bs99}.  With the
code having passed these tests, we are now implementing both
collisionless matter, which will be described elsewhere, and
hydrodynamics, as described below, to evolve the matter
self-consistently with the fields.

\section*{Relativistic Hydrodynamics}

For a perfect fluid, the stress-energy tensor can be written
\begin{equation}
T^{\alpha\beta} = (\rho_0 + \rho_0 \epsilon + P) 
	u^{\alpha} u^{\beta} + P g^{\alpha\beta},
\end{equation}
where $\rho_0$ is the rest mass density, $\epsilon$ the specific
internal density, $P$ the pressure, $u^{\alpha}$ the fluid four
velocity, and $g_{\alpha\beta}$ the four dimensional spacetime metric.
We construct constant entropy initial data with a polytropic equation
of state
\begin{equation}
P = K \rho_0^{\Gamma},
\end{equation}
where $\Gamma = 1 + 1/n$ and $n$ is the polytropic index, and where
we assume the polytropic constant $K$ to be unity without loss of
generality.  During the evolution, we adopt the gamma-law relation
appropriate for adiabatic flow,
\begin{equation}
P = (\Gamma - 1) \rho_0 \epsilon.
\end{equation}

Following~\cite{twb:s99,twb:sbs98}, we write the
equation of continuity
\begin{equation}
(\rho_0 u^{\alpha})_{;\alpha} = 0
\end{equation}
and the equations of motion
\begin{equation}
T^{\alpha\beta}_{~~~;\beta} = 0
\end{equation}
in the form
\begin{eqnarray}
\rho_{*,t} + (\rho_* v^i)_{,i} & = & 0, \label{E:twb:1}\\
e_{*,t} + (e_* v^i)_{,i} & = & 0,  \\
(\rho_* \tilde u_i)_{,t} + (\rho_* \tilde u_i v^j)_{,j} & = &
	- \alpha e^{6 \phi} P_{,i} - \alpha \rho_* \tilde u^0 \alpha_{,i} +
	\rho_* \tilde u_l \beta^l_{~,i} + \nonumber \\
	& &  \frac{\rho_* \tilde u_l \tilde u_m}{e^{4 \phi} \tilde u^0 } 
	\left( 2 \tilde \gamma^{lm}\phi_{,i} + 
	\frac{1}{2} \left(\tilde \Gamma^l_{ki}\tilde \gamma^{km}
	+ \tilde \Gamma^m_{ki}\tilde \gamma^{kl} \right) \right).
	 \label{E:twb:2}
\end{eqnarray}
Here we have defined the auxiliary quantities
\begin{eqnarray}
\rho_* & \equiv & \alpha e^{6 \phi} u^0 \rho_0,\\
e_* & \equiv &  \alpha e^{6 \phi} u^0 (\rho_0 \epsilon)^{1/\Gamma},\\
\tilde u_i & \equiv & (1 + \Gamma \epsilon) u_i, \\
\tilde u^0 & \equiv & (1 + \Gamma \epsilon) u^0, \\
v^i & \equiv & u^i/u^0 = - \beta^i + \gamma^{ij} u_j/u^0.
\end{eqnarray}
Similar equations have been used by many other groups
(e.g.~\cite{twb:hsw84} and references therein).  We integrate
equations~(\ref{E:twb:1})~-~(\ref{E:twb:2}) with an artificial
viscosity scheme suggested in~\cite{twb:on89} (see~\cite{twb:fmst98}
for implementations of more elaborate shock capturing schemes).  Given
$\rho_*$, $e_*$ and $\tilde u_i$ on a new timelevel, $u^0$ can be
found iteratively from the normalization relation 
$u^{\alpha} u_{\alpha} = -1$,
which yields
\begin{equation}
(\alpha u^0)^2 = 1 + 
	\frac{\tilde \gamma^{ij} \tilde u_i \tilde u_j}{e^{4 \phi}}
	\left( 1 + \Gamma 
	\frac{e_*^{\Gamma}}{\rho_* (\alpha u^0 e^{6 \phi})^{\Gamma - 1}}
	\right)^{-2}.
\end{equation}
The matter sources for the right hand sides of Einstein's equations
can then be constructed from these variables.

\section*{Numerical Results}

\begin{figure}[t!] 
\centerline{ \epsfig{file=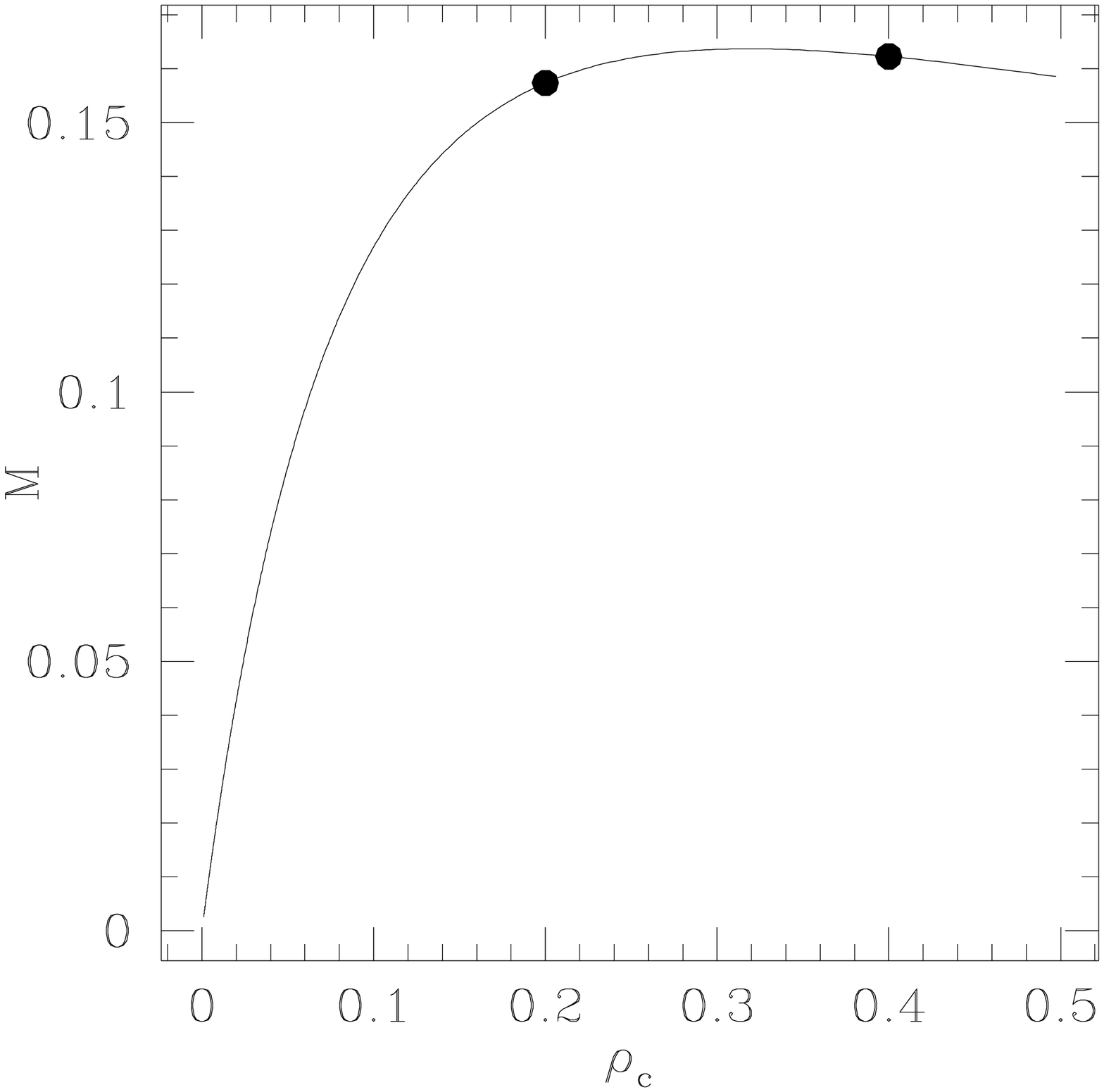,height=2.5in,width=2.5in}
	\epsfig{file=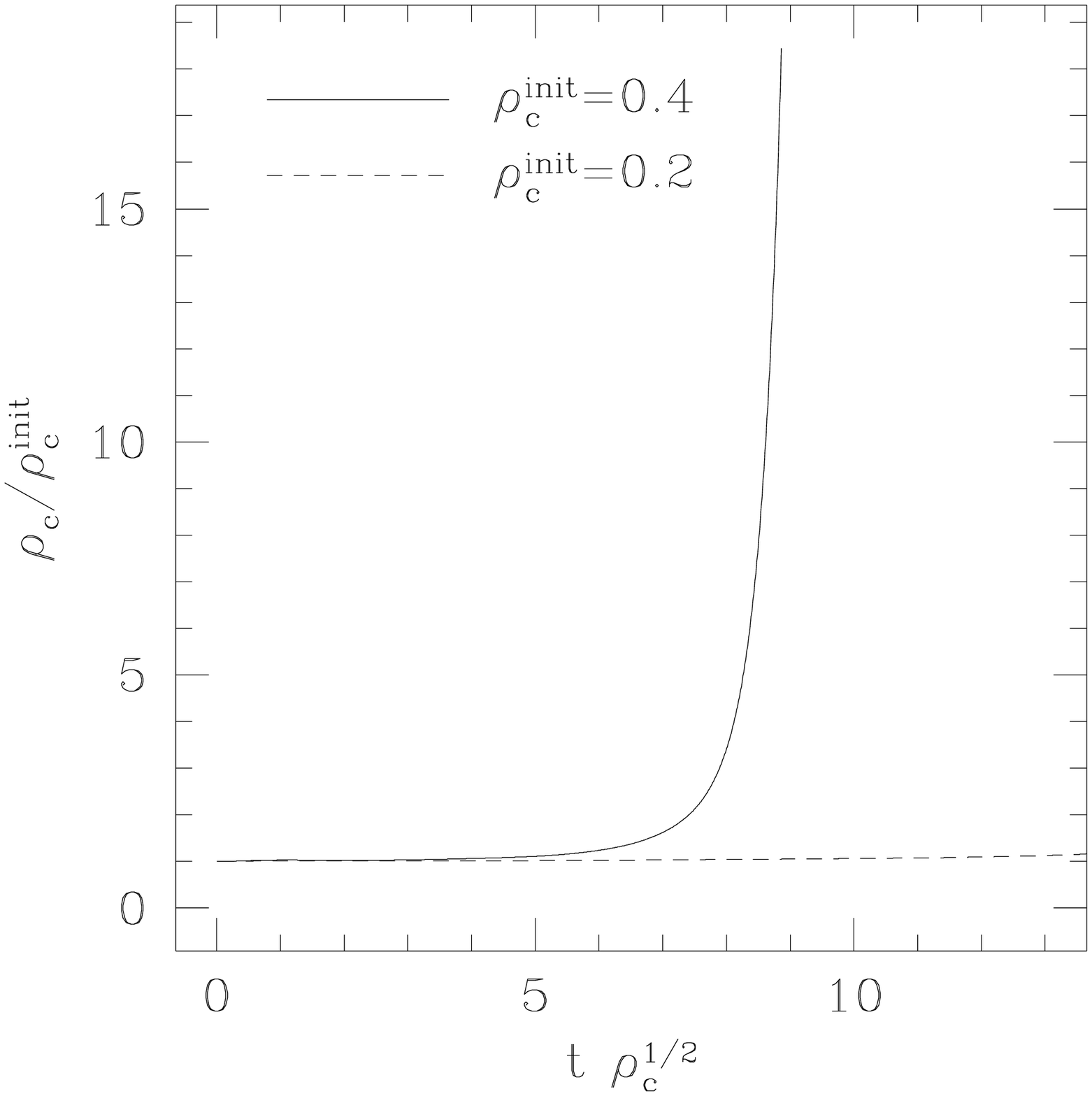,height=2.5in,width=2.5in} }
\caption{Mass versus central density for a $n=1$ polytrope (left
panel).  We chose the two configurations marked by the filled circles
($\rho_c = 0.2$ and $\rho_c = 0.4$) as initial data for our dynamical
simulations.  The central density as a function of time for the
evolution of the two initial data sets (right panel).  As expected,
the $\rho_c = 0.4$ configuration soon collapses, whereas the $\rho_c =
0.2$ configuration remains stable for several dynamical timescales.}
\vspace*{10pt}
\end{figure}

As a first test of our implementation of hydrodynamics, we adopt
the Oppenheimer-Volkov solution describing equilibrium neutron stars in
spherical symmetry as initial data and evolve these dynamically.
Constructing a sequence of such Oppenheimer-Volkov solutions for
increasing central rest mass densities $\rho_c$, the ADM mass $M$ of
the star takes a maximum $M_{\rm max}$ at $\rho_c^{\rm crit}$ (see
the left panel in Figure~1).  For central densities smaller than
$\rho_c^{\rm crit}$, the star is in stable equilibrium, while for
$\rho_c > \rho_c^{\rm crit}$ it is unstable and will collapse to a
black hole.

We adopt a polytropic equation of state with $\Gamma = 2$ ($n = 1$),
for which $\rho_c^{\rm crit} = 0.32$ and $M_{\rm max} = 0.164$ when $K
= 1$.  We choose as initial data the two configurations marked with
filled circles in Figure~1; a stable configuration with $\rho_c =
0.2$, $M = 0.157$ and $R/M = 5.5$, and an unstable configuration with
$\rho_c = 0.4$, $M = 0.162$ and $R/M = 4.4$.

In the right panel of Figure~1, we show the central density as a
function of time for the two initial configurations.  We performed
these runs on quite modest numerical grids with $(64)^3$ gridpoints
using cartesian coordinates and imposing outgoing wave boundary
conditions at $x,y,z = 2$.  We employ harmonic slicing and zero shift.
As expected, the unstable configuration soon collapses, while the
stable configuration remains stable for several dynamical
timescales (the period of the fundamental radial oscillation for a
$\rho_c = 0.2$ configuration is approximately $7 \rho_c^{-1/2}$,
compare~\cite{twb:s99}).  Ultimately, accumulation of numerical error
causes this configuration to collapse too, but this can be
delayed by increasing the grid resolution.

In Figure~2 we show density contours at the beginning and towards the
end of the simulation for the unstable configuration with $\rho_c =
0.4$ initially.  We also include arrows indicating the fluid flow
$\tilde u_i$.  The star is rapidly contracting and collapsing to a
black hole.  Up to the late stage of the collapse, the mass $M$ is
conserved to about 5 \%.  We can follow the collapse to about a 18
fold increase of the central density.  By this time, the central lapse
has decreased from 0.4 initially to about 0.03.

\begin{figure}[t!] 
\centerline{
	\epsfig{file=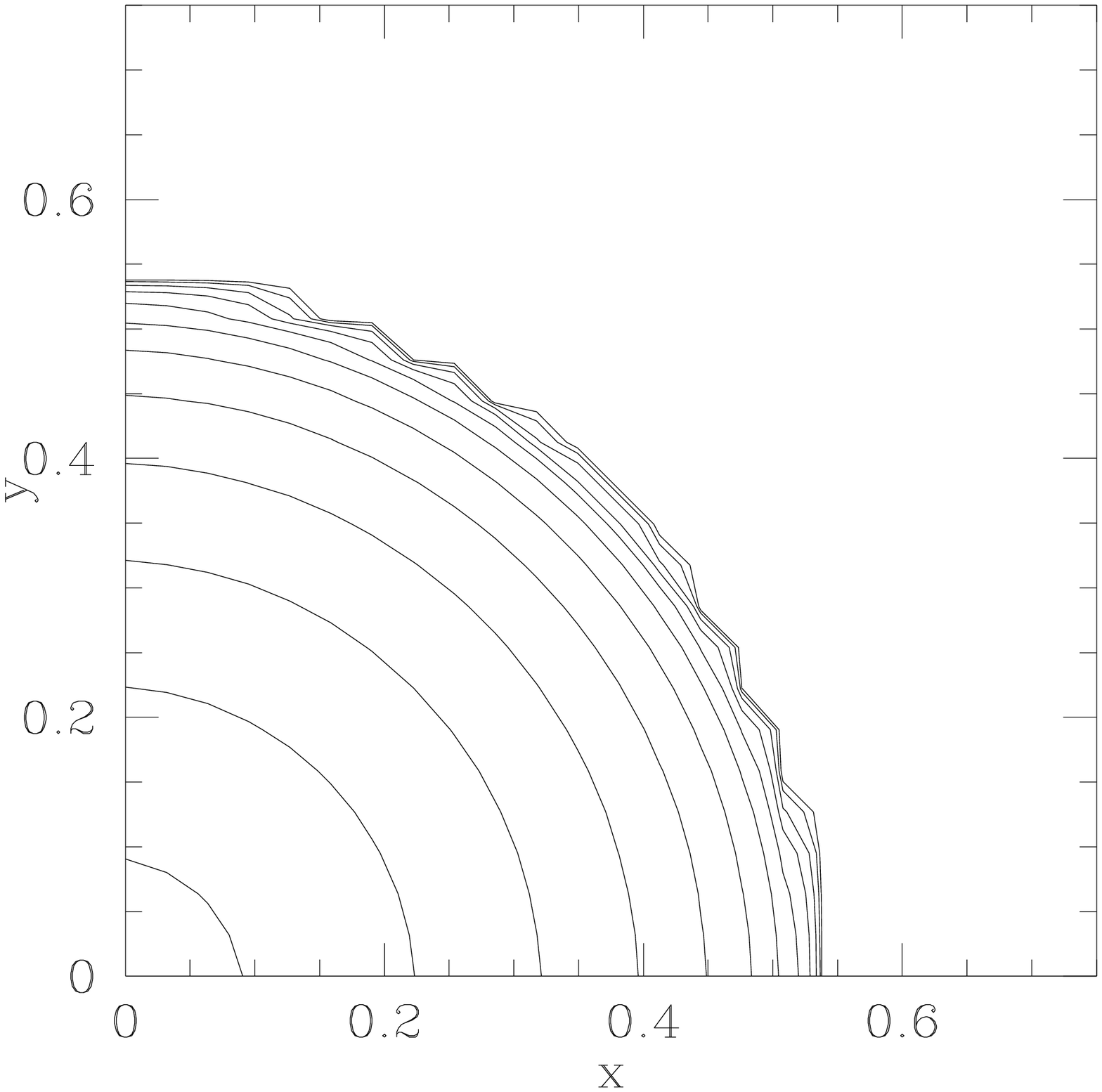,height=2.5in,width=2.5in}
	\epsfig{file=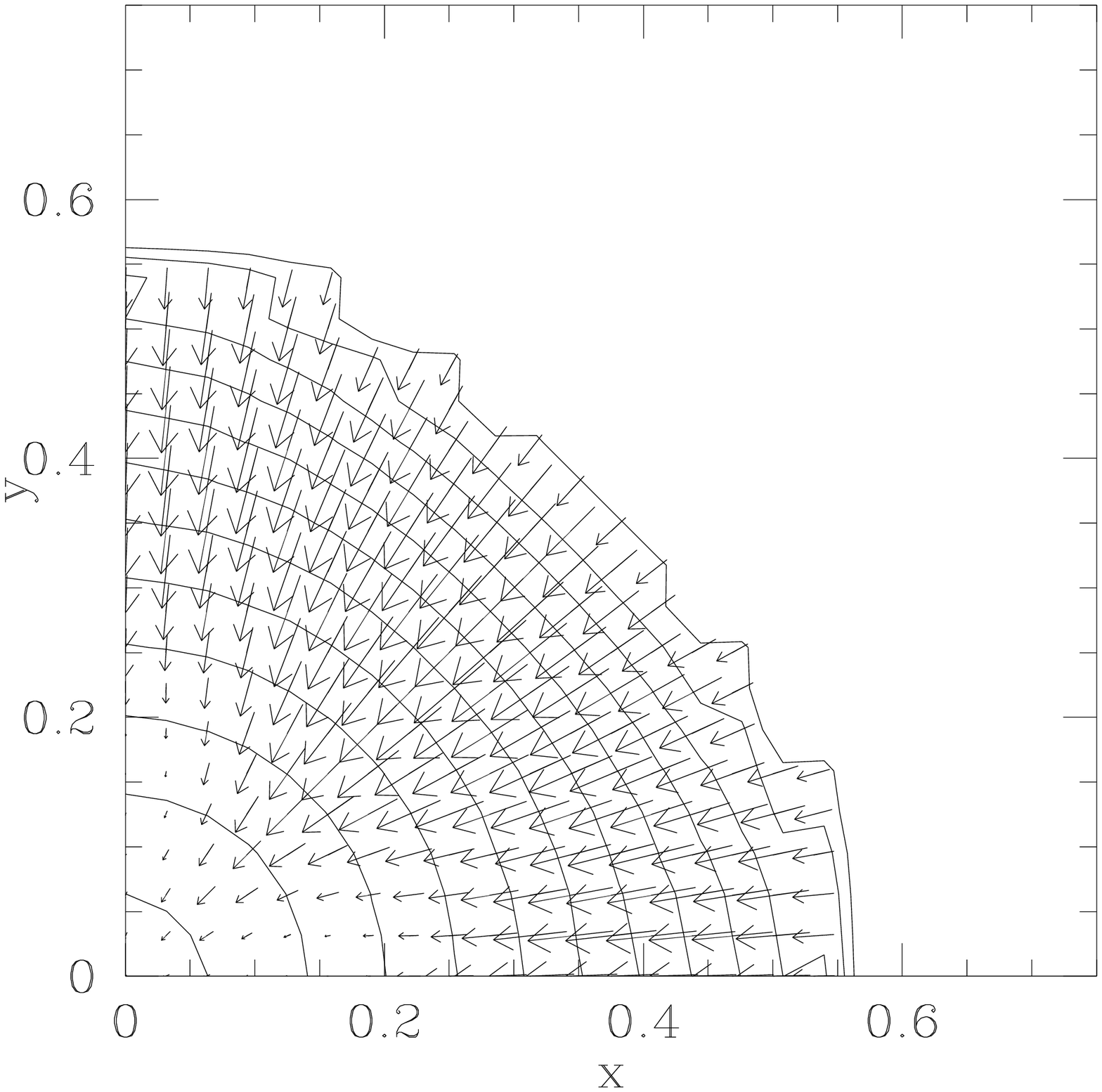,height=2.5in,width=2.5in}
	}
\caption{Rest mass density contours at $t = 0$ (left panel) and $t =
	13$ (right panel) in the $z = 0$ (equatorial) plane for the
	unstable configuration with $\rho_c = 0.4$ initially.  The
	contours logarithmically span densities between $\rho_c$ and
	$10^{-3} \rho_c$.  We also include arrows indicating the fluid
	flow $\tilde u_i$.}
\vspace*{10pt}
\end{figure}

\section*{Summary and Discussion}

We report on our systematic approach towards constructing a
fully relativistic hydrodynamics code in three spatial dimensions.  As
part of this program, we have developed a new formulation of
Einstein's equations which in several tests and applications has
proved to be much more suitable for numerical implementations than the
traditional ADM formulation~\cite{twb:bs99,twb:aablsst99}.
Mathematical properties of this formulation have been analyzed
in~\cite{twb:fr99}.  We have studied the evolution of small amplitude
gravitational waves to test dynamical transverse fields and have
inserted analytical solutions as matter sources (hydro-without-hydro)
to test highly relativistic longitudinal fields.

We outline our implementation of the relativistic equations of
hydrodynamics and present preliminary test results for spherical
neutron stars in hydrostatic equilibrium (Oppenheimer-Volkov stars).
As expected, we find that stable stars remain stable for several
dynamical timescales, while unstable stars soon collapse to black
holes.  We conclude that our method seems like a very promising
approach towards simulating the final plunge and merger
of binary neutron stars.

On a more speculative note, we point out that fully self-consistent
hydrodynamics may not be a feasable approach towards simulating the
coalescence and gravitational wave emission from neutron stars in the
intermediate inspiral phase.  In this epoch, the stars are very close
and interact through a strongly relativistic tidal field, but reside
outside the innermost stable circular orbit and hence move on a nearly
circular orbit.  Simulating the slow inspiral would require evolving
the stars for hundreds or thousands of orbits, which is presently
impossible.  It may be possible, however, to insert known
quasi-equilibrium binary configurations
(e.g.~\cite{twb:bcsst97,twb:bgm99}) into the field evolution code to
get the transverse wave components approximately.  Decreasing the
orbital separation (and increasing the binding energy) at the rate
consistent with the computed outflow of gravitational-wave energy
would generate an approximate strong-field wave inspiral pattern.
Such a ``hydro-without-hydro'' calculation may yield an approximate
gravitational waveform from inspiraling neutron stars, without having
to couple the matter and field integrations.

Calculations were performed on SGI CRAY Origin2000 computer systems at
the National Center for Supercomputing Applications, University of
Illinois at Urbana-Champaign.  This work was supported by NSF Grants
AST 96-18524 and PHY 99-02833 and NASA Grant NAG 5-7152 at Illinois.

\end{document}